\documentclass[11pt,a4paper]{article}
\usepackage[dvipdfmx]{graphicx}
\usepackage{amsmath,amssymb,enumerate}
\usepackage{subfigure, url, ulem}
\usepackage{cite,booktabs}
\usepackage{comment}
\usepackage{graphicx,rotating,colortbl}
\usepackage{color}
\usepackage{color}
\usepackage{youngtab}

\def\beq{\begin{eqnarray}}
\def\eeq{\end{eqnarray}}

\def\SU{\mathop{\rm SU}}

\begin{document}
\baselineskip 0.7cm
\begin{titlepage}

%\begin{flushright}
%KEK-TH-****
%\end{flushright}
%\vskip 1.35cm
~
\vskip 1.7cm
\begin{center}
{\Huge \bf 
Composite Accidental Axions 
}

%\author{
%{\large  Michele Redi\footnote{michele.redi@fi.infn.it}}\\
%[10mm] \normalsize\itshape INFN, Sezione di Firenze, Via G. Sansone, 1; I-50019 Sesto Fiorentino, Italy
%\\}

\vskip 1.2cm
Michele Redi$^a$, Ryosuke Sato$^{b,c}$,
\vskip 0.4cm

$^a${\it
INFN, Sezione di Firenze, Via G. Sansone, 1; I-50019 Sesto Fiorentino, Italy\\
}
$^b${\it
Department of Particle Physics and Astrophysics,\\
Weizmann Institute of Science, Rehovot 7610001, Israel\\
}
$^c${\it
Institute of Particle and Nuclear Studies,
High Energy Accelerator\\ Research Organization (KEK),
Tsukuba 305-0801, Japan
}

\vskip 1.5cm

\abstract{
We present several models where the QCD axion arises accidentally. 
Confining gauge theories can generate axion candidates whose properties are 
uniquely determined by the quantum numbers of the new fermions under the Standard Model. 
The Peccei-Quinn symmetry can emerge accidentally if the gauge theory is chiral. We generalise previous constructions in a unified
framework. In some cases these models can be understood as the deconstruction
of 5-dimensional gauge theories where the Peccei-Quinn symmetry is protected by
locality but more general constructions are possible.}
\end{center}
\end{titlepage}
\setcounter{page}{2}

%%%%%%%%%%%%%%%%%%%%%%%%%%%%%%%%%%%%%%%%%%%%%%%%%%%%%%%%
% intro
%%%%%%%%%%%%%%%%%%%%%%%%%%%%%%%%%%%%%%%%%%%%%%%%%%%%%%%%
\section{Introduction}

A great lesson of the Standard Model (SM) is the existence of accidental global symmetries that explain basic phenomenological features, for example the stability of the proton. 
Arguably the most attractive solution of the strong CP problem is provided by axions, Goldstone bosons (GB) of a $U(1)$ global symmetry, known as Peccei-Quinn (PQ) symmetry, dominantly broken by QCD anomalies \cite{Peccei:1977hh}.
Judging from the standpoint of the SM an important weakness of most axion models is that the global symmetry must be 
enforced by hand\footnote{The same criticism can be made for dark matter models where the cosmological stability is typically achieved invoking  discrete or continuous global symmetries. 
Models where dark matter is accidentally stable can be constructed using a new gauge theory, see \cite{strongDM}.}.
The quality of the PQ symmetry must be exceptional since even tiny explicit symmetry breaking effects would  
produce an axion potential incompatible with the solution of the strong CP problem \cite{Georgi:1981pu, Dobrescu:1996jp}.
If for example the PQ symmetry is broken by an operator of dimension $d$ suppressed by powers of the Planck scale, generically  
this induces and effective $\theta$-angle,
\begin{align}
\bar\theta \sim \frac 1 {\Lambda_{QCD}^4} \frac {\langle {\cal O}_d \rangle}{M_p^{d-4}}\,.
\end{align}
To explain the absence of neutron electric dipole moment, PQ breaking terms up to at least dimension 11 should be forbidden,
if the VEV is order $f_{\rm PQ}=10^{12}$ GeV. We regard this as a strong motivation to look for models where 
the PQ symmetry emerges as an accidental symmetry.

In this note we will construct very simple models where the axion emerges accidentally from the dynamics of 4 dimensional gauge theories.
We begin by observing that the only way to obtain accidental global symmetries, at least in weakly coupled 4D field theories, is in theories with  massless spin-1 particles. Since these are described by a gauge theory, the renormalizable Lagrangian enjoys an accidental global (flavor) symmetry when reducible representations of the gauge group are present. For the axion solution of the strong CP problem we need a $U(1)$ global symmetry anomalous under QCD. If the fermions are in a real representation of the gauge group (including QCD), mass terms can be written down and all global symmetries are not anomalous, as they are vectorial.  Therefore we are led to consider chiral gauge theories. In this case there are accidental  symmetries that are anomalous in the background of SM gauge fields.  Next the accidental symmetry should be spontaneously broken by the vacuum, producing the axion particle\footnote{If the vacuum does not break  the symmetry the $\theta-$angle can still be removed. Since the vacuum does not break color by anomaly matching this would imply the existence of colored massless fermions, which is excluded experimentally.}. 

A natural setting to construct chiral theories with the required properties is provided by Georgi's moose theories \cite{Georgi:1985hf}
with fermions that are bi-fundamentals fields of nearest neighbour gauge groups.
Such theories can also be understood as the deconstruction of an extra-dimension \cite{ArkaniHamed:2001ca}.
An axion model in the deconstruction framework can be found in Ref.~\cite{Hill:2002kq},
and axion models have  been constructed in extra-dimensions  \cite{Cheng:2001ys, Izawa:2002qk, Choi:2003wr, Grzadkowski:2007xm}. More possibilities exist in 4D. 
For example in \cite{Randall:1992ut} a model was presented where suppression of higher dimensional 
operators was achieved with a chiral gauging of the global symmetries. We extend these works in several directions and show the general 
conditions to construct composite accidental axion scenarios where operators up to dimension 11 or larger are forbidden.

The paper is organised as follows: in section \ref{sec:compositeaxion} we review and generalize  composite axion models based on $SU(N_c)$ gauge dynamics.
In section \ref{sec:accidentalPQ} we show how to realize the PQ symmetry as accidental symmetry of the dynamics. 
Several models are constructed that are distinguished by the way the SM gauge symmetry into the global symmetry of the dynamics.
Some phenomenological highlights are given in \ref{sec:phenomenology}. Conclusion are in section \ref{sec:conclusions}
and generalization to $SO(N_c)$ gauge theories can be found in the appendix.

\section{Composite Axions in vector-like models}
\label{sec:compositeaxion}

Axion-like particles are a generic prediction of confining gauge theories.
The first composite axion model was built by Choi and Kim (CK) in \cite{Kim:1984pt}. They introduced massless Dirac fermions with quantum numbers of
a color triplet and a singlet, charged under a new $SU(N_c)$ confining gauge interaction. In Weyl notation the fermions are in the representation,
\begin{equation}
(N_c,3)+(\bar{N_c},\bar{3})+(N_c,1)+(\bar{N_c},1)
\label{eq:CK}
\end{equation}
The dynamics of this theory is well known from QCD. Assuming that the $SU(N_c)$ gauge coupling is the strongest interaction its confinement at the scale $f$
spontaneously breaks the flavour symmetry $SU(4)_L\times SU(4)_R$ to $SU(4)_V$,  where the $SU(3)_c\subset SU(4)_V$  is gauged by QCD. 
This gauging leaves an exact singlet GB. To see this note that the 15 GBs in the adjoint representation of $SU(4)$ decompose under $SU(3)_c$ as,
\begin{equation}
15 = 8 \oplus 3 \oplus \bar 3 \oplus 1
\label{eq:CKGB}
\end{equation}
Color interactions explicitly break the global symmetry and colored GBs acquire mass from (calculable) loop corrections.
The chiral symmetries under which the singlet shifts instead is not broken so it remains exactly massless up to non perturbative effects induced by the anomaly. 
The singlet is associated to the combination of charges of the 8 Weyl fermions in eq. (\ref{eq:CK}),
\begin{equation}
T_{PQ}=({\rm Id}_{3}, {\rm Id}_{3},-3,-3)
\label{eq:genCK}
\end{equation}
This symmetry is anomalous in the background of QCD. This generates the term in the axion effective action,
\begin{equation}
-\frac {2 N_c} {32 \pi^2}\frac a {f_{PQ}} G_{\mu\nu} \tilde{G}^{\mu\nu}
\end{equation}
where $f_{PQ}$ is the decay constant of the axion precisely defined below. The dynamics of $a$ is then determined by QCD non-perturbative effects that drive
$a$ to 0 thus solving the strong CP problem of QCD. The orthogonal combination of charges $T_{\eta'}=(1,1,1,1)$ is instead anomalous under $SU(N_c)$. 
The corresponding GB is the analog of the $\eta'$ in QCD and acquires a mass proportional to $f_{PQ}$ from the $SU(N_c)$ anomaly.  

This construction can be easily generalized to arbitrary representations: for any choice of SM quantum numbers one can construct a composite axion model along the lines above.  
The flavor symmetry is  $SU(N_F)\times SU(N_F)$ with,
\begin{equation}
N_F= \sum_{i=1}^{N_S} d_i
\end{equation}
where $d_i$ is the dimension of the $i$-th SM representation and $N_S$ is the number of species.  
It is easy to see that the gauging of SM symmetries leaves $N_S-1$ exact GBs
(in addition the combination proportional to the identity is again anomalous under the $SU(N_c)$ gauge fields) 
that are singlets under the gauge interactions\footnote{If a SM rep appears with a multiplicity $\kappa$ a non Abelian chiral symmetry $U(\kappa)$ is preserved 
so that the number of singlets is increased to $\kappa^2$.}. If at least one fermion carries color charge one combination of singlets will be anomalous under QCD  
while the orthogonal $N_S-2$ combinations will have only anomalies with electro-weak gauge bosons and therefore remain massless until explicit breaking effects are included.

\begin{table}
\begin{center}
\begin{tabular}{c||c|c|c|c|c|c|c||c|c|c|c|c}
& $q_1$\,&  $q_2$\,& $N$\,  & $E/N$  \,\\  \hline
$D+L$		& $2$ & $-3$ & $4 N_c $ &  $-7/3$  \\ 
$Q+E$		& $1$ & $-6$ & $4 N_c$  &  $-13/3$ \\ 
$U+E$		& $1$ & $-3$ & $2 N_c$  &  $-10/3$ \\ 
$U+V$		& $1$ & $-1$ & $2 N_c$ &  $-4/3$ \\ 
$D+N$		& $1$ & $-3$ & $2 N_c$  &  $ 2/3$  \\ 
\end{tabular}
\end{center}
\caption{Anomaly coefficients for the axion singlet in $SU(N_c)$ models with various choices of reps of SM fermions. 
We use the notation as in~\cite{strongDM}.}
\label{table:CKexamples}
\end{table}

More in detail assigning a chiral charge $q_i$ to the i-th SM rep 
the combinations of charges that is not anomalous under $SU(N_c)$ and QCD satisfies
\begin{equation}
\sum_{i=1}^{N_S} q_i d_i =0\,,~~~~~~~~~~~~~~~~~~ \sum_{i=1}^{N_S} q_i N_i=0
\end{equation}
where $N_i$ is the color anomaly coefficient given in table \ref{tab:unificaxion} for branches of $SU(5)$ reps.
The QCD axion can be identified with the combination of charges $T_{PQ}$ orthogonal to these massless modes and to the identity (up to negligible corrections of order
$\Lambda_{QCD}^2/ \Lambda_{SU(N_c)}^2$). It is interesting to note that this type of models unavoidably not only solves the strong CP problem 
but also eliminates the  $SU(N_c)$ $\theta-$angle.
 
Once we identify the combination that corresponds to the QCD axion, anomalies are determined purely from group theory. 
Parametrizing the axion field as $\exp(i a \,T_{PQ}/f_{PQ})$ the low energy Lagrangian is\footnote{We normalize $T_{PQ}$ so that the periodicity of $a$ is a multiple of $2\pi f_{PQ}$. 
The axion decay constant is related to the decay constant of the $\sigma-$model by  $f_{PQ}^2=2 \sum_i d_i q_i^2 f^2$.},
\begin{equation}
{\cal L}_a=  \frac 1 2 (\partial a)^2-\frac 1 {32\pi^2}\frac a {f_{PQ}}\left[e^2 E\, F_{\mu\nu}\tilde{F}^{\mu\nu}+g_3^2\,N\, G_{\mu\nu}\tilde{G}^{\mu\nu}\right]
\end{equation}
where
\begin{equation}
E=4N_c\,{\rm Tr} (T_{PQ} Q^2)\  ,\qquad \hbox{and}\qquad 
N\delta^{AB}= 4 N_c\, {\rm Tr} \,(T_{PQ} T^A T^B).
\end{equation}
Here $Q$ is the electric charge matrix, $T^A$ are the $\SU(3)_c$ generators, and $T_{PQ}$ is the generator of the chiral symmetry of the axion. 
Note that because SM fermions have no PQ charge there is no UV contribution to the coupling to SM fermions $(\partial_\mu a)\bar{f}\gamma^\mu \gamma^5 f$.
As a consequence the couplings to matter are  model independent, see section \ref{sec:phenomenology}.

The structure of the low energy Lagrangian is fixed by the global structure of the vacuum and can be encoded in the Wess-Zumino-Witten term 
of the GBs effective Lagrangian.  In particular the ratio of anomalies $E/N$ that determines the coupling to photons  does not depend on the UV completion. 
For example in the case of two species we find,
\begin{equation}
\frac E N= \frac {d_2 E_1-d_1 E_2}{d_2 N_1-d_1 N_2}
\end{equation}
where $E_{1,2}$ and $N_{1,2}$ are the electromagnetic and color anomalies associated to each multiplet. A sample of models and their anomaly coefficients 
is given in table \ref{table:CKexamples}. 

\section{Accidental Peccei-Quinn symmetry}
\label{sec:accidentalPQ}

The composite axion models described above are vector-like so that PQ symmetry could be already violated 
by fermion mass terms. Moreover the higher dimensional operators that break the PQ symmetry should also be forbidden. 

Here we explore the possibility that the PQ symmetry of composite axion models is accidental.
We focus on the symmetry breaking pattern $SU(N_F)_L \times SU(N_F)_R \to SU(N_F)_V$ obtained with QCD-like dynamics
where the axion $a$ is identified with one of the GBs (extension to $SO(N)$ gauge theories is described in the appendix). 
In general, we can weakly gauge any subgroup $H_L \times H_R$  of the global symmetry group $SU(N_F)_L \times SU(N_F)_R$ as long as cancellation of gauge anomalies is satisfied. 
Because of the spontaneous symmetry breaking, the unbroken gauge group will be given by the intersection  of $H_L \times H_R$ with 
$\subset SU(N_F)_V$. In order to generate an anomaly with QCD the unbroken group should contain $SU(3)_c$. There are two possibilities:
\begin{itemize}
\item $SU(3)_c \subset H_V$: this gauging is not chiral so that anomaly cancellations is automatic. 
The gauge theory allows mass terms that would break explicitly the PQ symmetry. 
\item $SU(3) \subset H_R$: this gauging is chiral and $SU(3)_c$ at low energy emerges as the unbroken diagonal subgroup. 
Mass terms are forbidden.
\end{itemize}

In what follows we explain in detail how render the PQ symmetry accidental to the required accuracy.

\subsection{$SU(3)_c \subset H_V$}\label{sec:moose}

\begin{figure}[t]
\centering
\includegraphics[width=0.8\hsize]{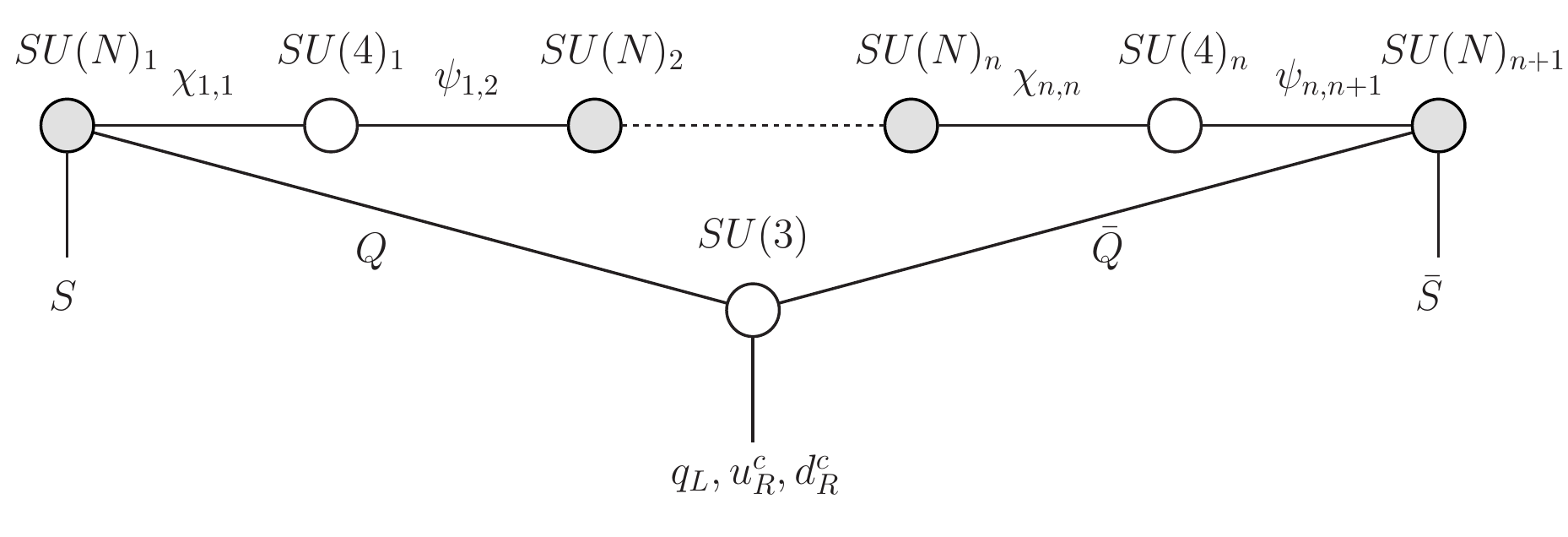}
\caption{Moose diagram of accidental axions based on the pattern $SU(4)_L\times SU(4)_R/SU(4)$. $Q$ transforms as SM color triplet and $S$ as a singlet.}\label{fig:moose}
\end{figure}

This is the same embedding as the CK model.
PQ breaking mass terms and higher dimensional operators  can be forbidden making the theory chiral. 
This can be achieved introducing gauge fields with fermions in a bi-fundamental representations 
under nearest neighbour gauge groups \cite{Georgi:1985hf}, as described by the moose in Fig.~\ref{fig:moose}.

The moose describes a gauge theory $ \prod_{i=1}^{n+1} SU(N_c)_i \times \prod_{j=1}^n SU(M)_j$ 
with Weyl fermions transforming as bi-fundamentals under nearest neighbour gauge fields.
Explicitly we take Weyl fermions $\chi_{i,i}$, $\psi_{i,i+1}$ transforming as $(N_c,\bar M,1)$ and $(1,M,\bar N_c)$
under $SU(N_c)_i \times SU(M)_i \times SU(N_c)_{i+1}$ gauge symmetry. The theory so constructed is chiral. No mass terms can be written down
and the global symmetry is then accidentally $SU(N_F)\times SU(N_F)$. 

The dynamics of the theory is as follows. We assume that $SU(N_c)$ gauge groups confine at scale $f_i$.
This spontaneously breaks the global symmetries that are (weakly) gauged by $SU(M)$ gauge fields. One obtains
a theory with a global symmetry  $SU(M)_L\times SU(M)_R/SU(M)_V$ broken at the scale,
\begin{equation}
\frac 1 {f^2}= \sum_{i=1}^{n} \frac 1 {f_i^2}
\end{equation}

Starting from this chiral moose we can proceed exactly as in the previous section gauging a subgroup of the unbroken global group
that corresponds to SM symmetries. The anomalies are identical to that case being determined by the number of colors
of the boundary confining gauge theories. Explicitly for the CK model $Q$ and $S$ transform as $(3,\bar N_c)$ and $(1,\bar N_c)$ under $SU(3)_c \times SU(N_c)_1$,
and $\bar Q$ and $\bar S$ behave as $(\bar 3,N_c)$ and $(1,N_c)$ under $SU(3)_c \times SU(N_c)_{n+1}$.
Simplest gauge invariant PQ breaking operators are obtained by stringing together the fermions along the moose, 
i.e. $Q( \Pi_{i=1}^n \chi_{i,i}\psi_{i,i+1})\bar{Q}$ whose dimension is $3n+3$.

In general suppression of higher dimensional operators in controlled by the ``length'' of the moose, i.e. how many sites it contains.
It is interesting to note that there is a limit to the possible suppression. Indeed the low energy SM couplings are given by,
\begin{equation}
\frac 1 {g_{SM}^2}=\frac 1 {g_0^2} +\sum_{i=1}^n \frac 1 {g_i^2}
\end{equation}
so that perturbativity of gauge couplings implies the mild constraint $n\ll 16 \pi^2/g_{SM}^2$.

All this mimics the locality of an extra-dimension and in fact it is just the deconstruction of a 5D gauge theory. 
It is interesting to consider the continuum limit of this theory, see \cite{adsqcd} for the same construction applied to QCD. 
This is given by an $SU(M)$  gauge theory  in a 5 dimensional interval with appropriate boundary conditions that reproduce the spontaneous breaking 
of the global symmetry.
As discussed in Ref.~\cite{Skiba:2002nx},
the effects of anomalies are encoded in the Chern-Simons term,
\begin{equation}
%S_{CS}= \frac {N_c}{24\pi^2} \int d^4x \int_0^L dz  \epsilon^{MNOPQ} {\rm Tr}\left[A( dA)^2+\frac 3 2 A^3(dA)+\frac 3 5 A^5\right]
S_{CS}= \frac {N_c}{96\pi^2} \int d^4x \int_0^L dz  \epsilon^{MNOPQ} {\rm Tr}\left[A_M G_{NO} G_{PQ}+ i A_M A_N A_O G_{PQ}-\frac 2 5 A_M A_N A_O A_P A_Q\right]
\end{equation}
In this construction the axion is the Wilson line $A_5^{PQ}$ and receives an anomalous coupling to gluons from the Chern-Simons term.
Strictly only certain 5D models can be reproduced with simple $SU(N_c)$ gauge dynamics.
For example the axion model in \cite{Choi:2003wr} considers an $SU(3)\times U(1)$ gauge theory in 5 dimensions. 
This can be thought as the truncation of the action above.

%\subsubsection{Accidental CK model}

%\begin{tabular}{|c||c|c|c|c|c|c|}
%\hline
%& $Q$ & $S$ & $\bar Q$ & $\bar S$ & $\chi_{i,i}$ & $\psi_{i,i+1}$ \\\hline\hline
%$U(1)'_B$ & 1 & 1 & $-1$ & $-1$ & $-1$ & 1 \\\hline
%$U(1)'_Y$ & 1/3 & $-1$ & $-1/3$ & 1 & 0 & 0 \\\hline
%$U(1)_{\rm PQ}$ & 1 & $-3$ & 1 & $-3$ & 0 & 0 \\\hline
%\end{tabular}

\subsubsection{Accidental symmetries}

The models presented above naturally feature accidental symmetries that are only broken by non renormalizable interactions.
Some of these symmetries (vectorial) are respected by the vacuum while others (chiral) are spontaneously broken.
The first kind leads to accidentally stable particles. These are:
\begin{itemize}
\item{{\bf Baryon number:} The rotation of all the Dirac fermions with an equal phase $\psi_i\to e^{i \alpha}\psi_i$  guarantees the stability of baryon states
$\epsilon^{i_1\dots i_N}\psi_{i_1}\dots \psi_{i_N}$.}
\item{{\bf Species number:} In models with more than one SM representation we can rotate two fermions independently. Pions made of different species carry a net 
species number and are therefore stable.}
\end{itemize}
The  stability of these states  leads to phenomenological problems unless the scale of inflation is below $f$, see section \ref{sec:phenomenology}. 

Chiral symmetries are responsible for the existence of composite axions. These are singlets that do not carry species number, indeed
they couple to gluons and photons. A model independent PQ breaking operator is the ``Wilson line'' operator such as $Q \chi_{1,1} \psi_{1,2} \cdots \chi_{n,n} \psi_{n,n+1} \bar Q$,
whose dimension is $3n+3$ where $n$ is the number of sites. For the axion to solve the strong CP problem one should have $n>3$ for $f_{PQ}= 10^{12}$ GeV.
The other terms that can break explicitly the global symmetries  are baryon operators\footnote{Since these operators do not have VEV bounds on their dimensionality will be weaker, see \cite{Dobrescu:1996jp}.}. From a fundamental representation we can build  a singlet of $SU(N_c)$ 
using the $\epsilon$ tensor with $N_c$ fermions. A singlet under the SM can be constructed multiplying together various of these terms. The lowest dimensional operator
that breaks the symmetry is model dependent. Consider for example the model $D+L$. The lowest dimensional baryon-like operator has the structure 
$D^{3 N_c} L^{2 N_c}$ which has sufficiently high dimensionality even for  $N_c=3$. In models with singlets lower dimensional PQ breaking 
operators exist. Indeed in this case $N^{N_c}$ is a singlet so that $N_c> 4 (8)$ for $N_c$ odd (even).

\subsection{$SU(3) \subset H_R$}
\label{sec:axialgauge}

In the construction above the SM gauges a subgroup of the unbroken global symmetry and chirality of the theory 
is induced by the moose structure. Another possibility that we now explore is that the SM 
gauge group arises as the unbroken group of a chiral gauging of the  global symmetry $SU(N_F)_L\times SU(N_F)_R/SU(N_F)_V$. 
As we will see this  further constrains the possible gauge invariant operators.

To construct a chiral composite axion model starting from $SU(N_c)$ gauge theory with vectorial fermions we need to  gauge a chiral subgroup of the global symmetry so that the unbroken 
group contains the SM and a spontaneously broken $U(1)$ symmetry that is anomalous under QCD. In general we can divide the generators into broken and unbroken ones that will decompose into representations of the SM symmetry. The gauging explicitly breaks some of the global symmetries; the broken generators that are gauged correspond to massive gauge bosons while 
the GBs whose shift symmetry is explicitly broken by the gauging acquire a mass. 

Since the independent gauging of $SU(N_F)_L$ and $SU(N_F)_R$ is chiral only certain subgroups can be gauged consistently. 
For example gauging the full $SU(N_F)_L\times SU(N_F)_R$ is anomalous. The effect of anomalies can be encoded in the Wess-Zumino-Witten term associated to the coset
and does not depend on the UV completion. The general condition for the gauging to be possible is that,
\begin{equation}
{\rm Tr}(T_L^3)= {\rm Tr}(T_R^3)
\label{eq:WZWanomaly}
\end{equation}
which is automatically satisfied when we gauge the unbroken group as in section \ref{sec:moose}.

\begin{figure}[t]
\centering
\includegraphics[width=0.7\hsize]{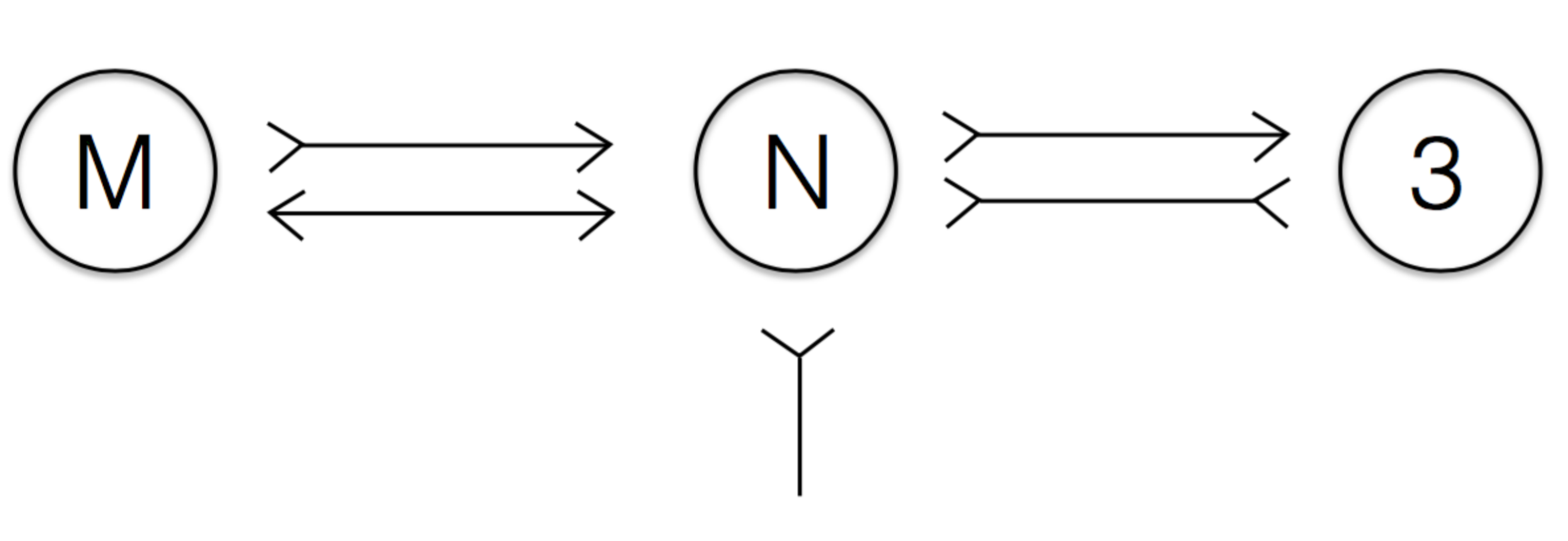}
\caption{
Moose diagram for axial models. Left and right arrow correspond to fundamental and anti-fundamental representations.
}\label{fig:axial}
\end{figure}

Another possibility that we pursue here is to perform a gauging so that left and right side of eq. (\ref{eq:WZWanomaly}) are individually zero.
The simplest choice that produces a composite axion model is:
\begin{equation}
(N_c,m,1)\oplus(\bar{N_c},1,\bar{3})\oplus(N_c,\bar{m},1)\oplus(\bar{N_c},1,3)\oplus 2(m-3)(\bar{N_c},1,1)
\end{equation} 
This corresponds to a theory with flavor symmetry $SU(2m)_L\times SU(2m)_R /SU(2m)_V$ where the gauged $SU(m)$ is embedded in  $SU(2m)_L$ so that the fundamental representation decompose as $2m= m +\bar{m}$ which is anomaly free and similarly $SU(3)$ in $SU(2m)_R$.  In the UV  the SM quarks are most simply charged under $SU(3)_R$ that is spontaneously broken by the strong dynamics. 

The chiral gauging of $SU(m)$ and $SU(3)$  breaks explicitly some 
of the symmetries leaving an exact,
\begin{equation}
\frac {SU(m)\times U(1) \times SU(3)\times SU(2m-6) \times U(1) \times U(1) \times U(1)}{SU(3)\times U(1) \times U(1)}\
\end{equation}
The unbroken $SU(3)$ can be identified with $SU(3)_c$ at low energy.  
$SU(m)_L$ is gauged and spontaneously broken so that the corresponding GBs are eaten while GBs charged under the SM 
acquire mass from gauge interactions. Singlets remain massless and a linear combination is anomalous under QCD playing the role
of the axion\footnote{The condensates are $(N_c,m,1)\times [(\bar{N_c},1,\bar{3})\oplus (m-3)(\bar{N_c},1,1)]\,,~~ (N_c,\bar{m},1)\otimes [(\bar{N_c},1,3)\oplus (m-3)(\bar{N_c},1,3)]$
and the singlets correspond to the $U(1)$ chiral rotations spontaneously broken by the condensate.}.

It is interesting to understand how the strong CP problem is solved from the point of view of the high energy theory.
The QCD $\theta-$angle originates from $SU(m)$ and $SU(3)$ $\theta-$angles. The anomalous $U(1)$'s allow to remove both of them. 
Let's consider the $U(1)'s$ under which the i-th representation has charge $q_i$. We have the following mixes anomalies,
\begin{eqnarray}
&&U(1)\times SU(3):\,~~~~~~~~~~~~~~~~~~~~~ N_c q_{\overline{N}\overline{3}}+N_c q_{\overline{N}3}\nonumber \\
&&U(1)\times SU(m):\,~~~~~~~~~~~~~~~~~~~~N_c q_{Nm}+N_c q_{N\bar{m}}\nonumber \\
&&U(1)\times SU(N_c):\,~~~~~~~~~~~~~~~~~~~~3 q_{\overline{N}\overline{3}}+3 q_{\overline{N}3}\nonumber +m q_{Nm}+m q_{N\bar{m}}+\sum_{i=1}^{2m-6} q_{\overline{N}^i}
\end{eqnarray}
For $m>3$ one can choose combinations that are anomalous under each gauge group removing all the $\theta$ angles in the theory.
For $m=3$ instead the combination anomalous under $SU(3)_c$ is also anomalous under $SU(N_c)$. Therefore the axion acquires mainly 
a potential from  $SU(N_c)$ and does not solve the $SU(3)_c$ strong CP problem.

We can derive the results  above in a slightly different way following Ref. \cite{thaler}. The gauging of $SU(m)$ on the left reduces the global symmetry to\footnote{An even simpler possibility is to gauge the full anomaly free $SO(2m)_L$ leaving a global symmetry $SU(2m)/SO(2m)$. The pattern so obtained is identical to the one obtained in $SO(N)$ gauge theories with vectorial fermions, see appendix.},
\begin{equation}
\frac {SU(2m)\times U(1)}{SU(m)\times U(1)}
\end{equation}
The adjoint representation of $SU(2m)$  decomposes under $SU(m)$ as,
\begin{equation}
{\rm Adj}_{SU(2m)}= 1 + \yng(1,1) +\overline{\yng(1,1)} + \yng(2) +\overline{ \yng(2) }+2 {\rm Adj}_{SU(m)}
\end{equation}
The gauging of $SU(3)$ on the right corresponds to a gauging of a subgroup of the unbroken group so that the corresponding gauge fields remain massless. 
We can decompose the GBs in $SU(3)$ reps.
For example for $m=4$ we find,
\begin{equation}
{\rm GBs}= 4\times 1  \oplus 4 \times 3 \oplus 4 \times \bar{3} \oplus 6 \oplus \bar{6} \oplus  8   
\end{equation}
The charged GBs  acquire mass from color interactions that explicitly break their chiral symmetry. Singlets instead 
remain exact GBs up to anomalies. The QCD axion is then identified as the combination of charges anomalous under $SU(3)$
while the orthogonal combinations remain exactly massless. This formulation makes it obvious to compute anomalies.
The singlets GBs corresponds to the traceless diagonal generators of $SU(2m)_R$. For the QCD axion combination the charges are identical
to the ones in section \ref{sec:compositeaxion} leading to the same anomaly coefficients.  Compared to those models we also obtain extra axion-like particles 
with no QCD anomalies.

In this model the combination of SM and $SU(N_c)$ gauge symmetries is chiral so that no mass terms can be written down. 
We can however write dimension-6 operators that break the PQ symmetry, namely,
\begin{equation}
[(N_c,m,1)(N_c,\bar{m},1)][(\bar{N_c},1,\bar{3})(\bar{N_c},1,3)]
\end{equation}
Also in this case one can associate the suppression of higher dimensional operators to the length of the moose diagram. 
In Fig. \ref{fig:axial} the length of the gauge invariant loop connecting all the sites is 4 so that mass terms are forbidden while 4-Fermi operators are allowed.
Higher dimension operators can be further eliminated replacing the $SU(N_c)$ gauge theory with a moose model as in the vectorial models.
Another possibility would be to consider a more complicated chiral gauging of the global symmetries.

Let us discuss the running of gauge couplings. For the confining group,
\begin{equation}
\frac {d} {dt}\frac 1 {\alpha_N}=\frac {1} {6\pi}(11 N_c- 4  m).
\end{equation}
Asymptotic freedom of $SU(N_c)$ requires,
\begin{equation}
m< \frac {11} 4 N_c
\end{equation}
An important constraint on this type of models arises  from the absence of Landau poles below the Planck scale. 
This is necessary to justify the suppression of higher dimensional operators by that scale. For $\alpha_s$ one finds,
\begin{equation}
\frac 1 {\alpha_s(M)}\approx 30 + \frac 7 {2\pi} \log \frac M {10^{12}\,{\rm GeV}}- \frac {N_c} {3\pi} \log \frac M {10^{12}\,{\rm GeV}} 
\end{equation}
Requiring $\alpha_s(M_p)< 1$ implies $N_c< 28$ which is a mild constraint.

\subsubsection{Example: $SU(8)/SO(8)$}
\label{sec:SU8SO8}

Consider the gauge theory $SU(N_c)\times SO(8)\times SU(3) \times U(1)$ with the following fermion representations,
\begin{equation}
r_L=(N_c,8,1)\,,~~r_{R_{3x}}=(\overline{N_c},1,3)_x\,,~~r_{R_{\overline{3x}}}=(\overline{N_c},1,\overline{3})_{-x}~~r_{R_{1y}}=(\overline{N_c},1,1)_y\,,~~r_{R_{\overline{1y}}}=(\overline{N_c},1,1)_{-y}
\end{equation}
The flavor symmetry of the $SU(N_c)$ gauge theory is $SU(8)_L\times SU(8)_R$. The gauging of $SO(8)_L$ leaves at low energies the pattern of
symmetry breaking,
\begin{equation}
\frac {SU(8)}{SO(8)}
\end{equation}
The 35 GB decompose under $SU(3)_c \times U(1)_Y$ contained in the unbroken group as,
\begin{equation}
{\rm GBs}= 3\times 1_0 \oplus 3_{x+y} \oplus {\bar{3}}_{-x-y} \oplus 3_{x-y} \oplus {\bar{3}}_{-x+y} \oplus 6_{2x} \oplus {\bar{6}}_{-2x} \oplus 8_0
\end{equation}
In the low energy theory the singlet anomalous under QCD corresponds to the $SU(8)_R$ generator $({\rm Id}_{3}, {\rm Id}_{3},-3,-3)$ as in eq. (\ref{eq:genCK}).
The anomalies are then identical to that case.

From the $SU(N_c)\times SO(8)\times SU(3) \times U(1)$ point of view this can be understood as follows: The condensates are given by,
\begin{equation}
r_L r_{R_{3x}}\,,~~~~~r_L r_{R_{\overline{3x}}}\,,~~~~~r_L r_{R_{\overline{3x}}}\,,~~~~~r_L r_{R_{1y}}\,,~~~~~r_L r_{R_{\overline{1y}}}
\end{equation}
The vacuum is invariant under the rotation,
\begin{equation}
T_1= (1,-1,-1,-1, -1)
\end{equation}
whose mixed anomalies  with $SU(N_c)\times SO(8)\times SU(3) \times U(1)$ are,
\begin{equation}
A_H=(0, 2 N_c, - 2 N_c)
\end{equation}
The orthogonal spontaneously broken symmetries and the corresponding anomalies are,
\begin{eqnarray}
&&T_2:~~~~~(0,1,1,-3,-3)\,,~~~~~~~~~~A_2= (0, 0, 2 N_c, 4 N_c(3 x^2 - 3 y^2))\nonumber\\
&&T_3:~~~~~(0,1,-1,0,0)\,,~~~~~~~~~~A_3=(0, 0, 0, 0)\nonumber \\
&&T_4:~~~~~(0,0,0,1,-1)\,,~~~~~~~~~~A_4= (0, 0, 0, 0)\nonumber\\
&& T_5:~~~~~(1,1,1,1,1, 1)\,,~~~~~~~~~~A_5=(8, N_c, 2 N_c ,2 N_c( 3 x^2 + y^2))
\end{eqnarray}
where the last generator corresponds to the overall anomalous $U(1)$.
The are three independent combinations that are anomalous under each gauge group. This allows to remove all the
$\theta$ angles from the theory. Note  that this requires both broken and unbroken symmetries.

%\subsubsection{Example: $SU(N_c)\times SU(4)\times SU(3)$}

%We now describe in detail the minimal model with $m=4$.
%Denote the reps as,
%\begin{eqnarray}
%&&r_1=(N,4,1)\,,~~~~~~r_2=(\overline{N},1,\overline{3})\,,~~~~~~~r_3=(\overline{N},1,1)\nonumber\\
%&&r_4=(N,\overline{4},1)\,,~~~~~~r_5=(\overline{N},1,3)\,,~~~~~~~r_6=(\overline{N},1,1)
%\end{eqnarray}
%The condensates are,
%\begin{equation}
%r_1 r_2\,,~~~~r_1 r_3\,,~~~~r_4 r_5\,,~~~~r_4 r_6
%\end{equation}
%The unbroken $U(1)$ symmetries correspond to the following combinations of charges,
%\begin{equation}
%U(1)_1:~~~(1,-1,-1,0,0,0)\,,~~~~~~~U(1)_2:~~~(0,0,0,1,-1,-1)
%\end{equation}
%The anomalies under $SU(N_c)\times SU(M)\times SU(3)$ are,
%\begin{equation}
%A_1=A_2 :~~~~~~ (0,N_c,-N_c)
%\end{equation}
%The 4 broken $U(1)'s$ and their anomalies are instead,
%\begin{eqnarray}
%&&U(1)_3:~~~~~(1,1,0,0,0,0)\,,~~~~~~~~~~A_3:~~~~~(7, N_c, N_c)\nonumber\\
%&&U(1)_4:~~~~~(1,-1,2,0,0,0)\,,~~~~~~~~~~A_4:~~~~~(3, N_c,- N_c)\nonumber \\
%&&U(1)_5:~~~~~(0,0,0,1,1,0)\,,~~~~~~~~~~A_5=A_3\nonumber\\
%&&U(1)_6:~~~~~(0,0,0,1,-1,2)\,,~~~~~~~~~~A_6=A_4
%\end{eqnarray}
%The are three independent combinations that are anomalous under each gauge group. This allows to remove all the
%$\theta$ angles from the theory. Note however that this requires both broken and unbroken symmetries (the latter is similar
%to baryon number that is anomalous in the SM).

%To compute the anomalies at low energy we note that the gauging 
%\begin{equation}
%{\rm GBs}= 4\times 1  + 4 \times 3 + 4 \times \bar{3}+ 6 +\bar{6} +  8   
%\end{equation}

\subsection{SM + Axial Gauging }
\label{sec:randall}

\begin{figure}[t]
\centering
\includegraphics[width=0.6\hsize]{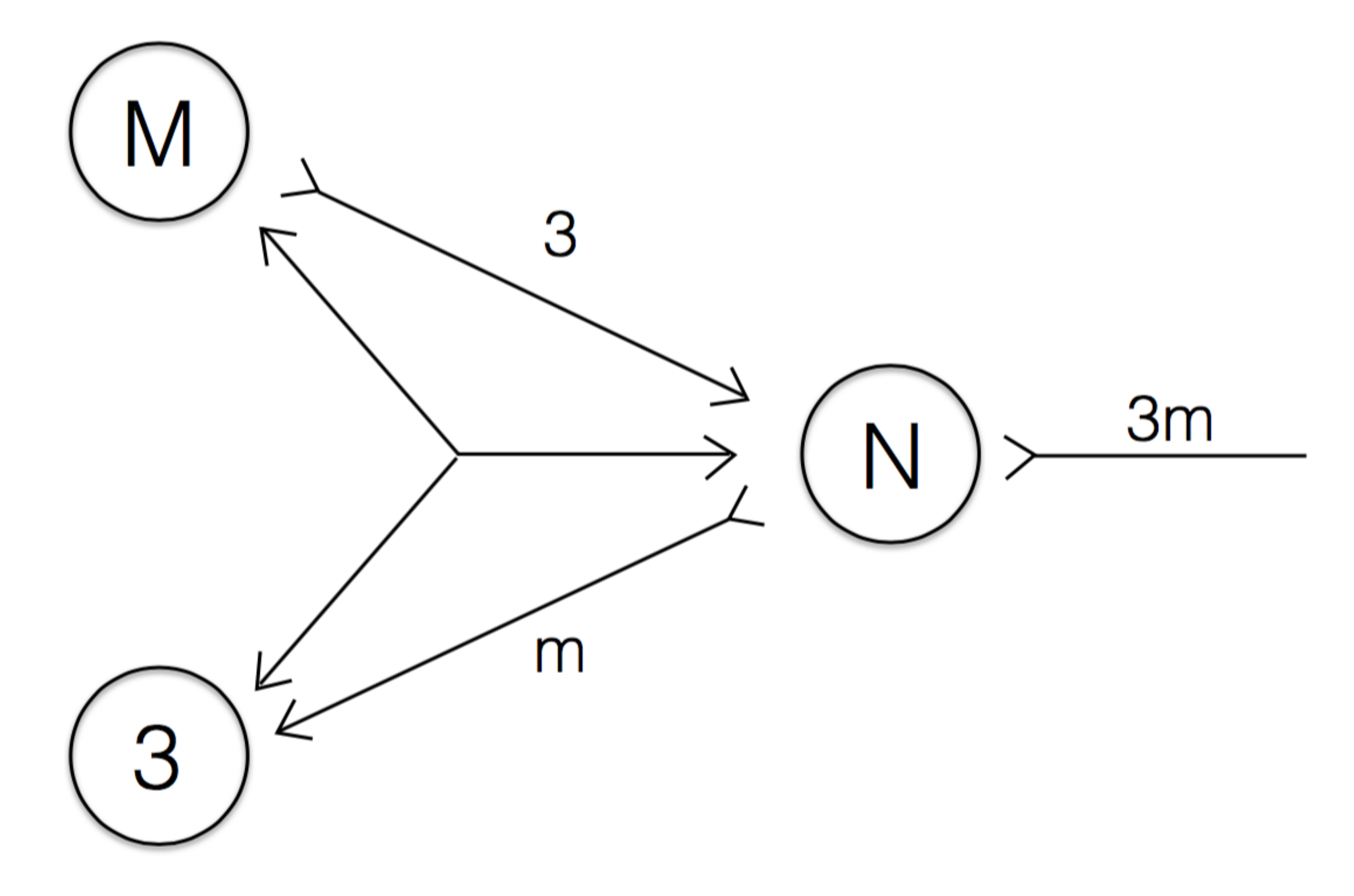}
\caption{Moose diagram of composite axion model of \cite{Randall:1992ut}.}
\label{fig:randall}
\end{figure}

Yet another generalization is provided by the model in \cite{Randall:1992ut}. 
Consider $SU(N_c)\times SU(m)\times SU(3)$ gauge theory. An anomaly free set of fermions is,
\begin{equation}
(N_c, m, 3)+ 3 (N_c, \bar{m},1)+ m (\bar{N_c}, 1,\bar{3})+3m (\bar{N_c},1,1)
\end{equation}
Among the GBs the one corresponding to a rotation of the colored fields and a rotation 
in the opposite direction of the $SU(3)_c$ singlets  remains exact and is anomalous under QCD providing a composite axion candidate. 

From a group theory perspective in this model the SM gauge symmetry is a subgroup of the unbroken group as in the CK model. 
A further chiral gauging, analogous to $SU(2)_L$ in the SM,  of the broken generators is performed so that the theory is chiral. 
From the point of view of the moose this corresponds  to introducing fields that are tri-fundamental as in Fig.~\ref{fig:randall}.

One interesting feature of this model is that the value of $m$ controls the dimension of the operators
that break the accidental axion global symmetry. The model however suffer from various constraints discussed in \cite{Dobrescu:1996jp}. 
There are $12 N_c m$ chiral fields charged under QCD. Asymptotic freedom limits how large $m$ and $N_c$ can be. 
Indeed requiring that operators that violate the global symmetry up to dimension 10 are forbidden only the values
$m=4$ and $N_c=5$ are allowed. 

To ameliorate the constraints on this model one could combine the gauge structure above with moose structure and obtain further suppression of higher dimensional operators. 
Also in this case  adding more sites to the moose does not change the global structure of the global symmetries. 

\section{Phenomenology}
\label{sec:phenomenology}

In this section we discuss various features typical of composite axion models.

\subsection{Coupling to Matter}

In the composite axion models studied here SM fields are not charged under the PQ symmetry. As a consequence 
the couplings to matter are model independent and can be computed with high precision using chiral Lagrangian methods \cite{diCortona:2015ldu}. 
In particular this implies the following couplings to nucleons,
\begin{equation}
-0.47(3) \frac N{2 f_{PQ}} \partial_\mu a \,\bar{p} \gamma^\mu \gamma^5 p-0.02(3) \frac N{2 f_{PQ}} \partial_\mu a\, \bar{n} \gamma^\mu \gamma^5 n
\end{equation}
Derivative couplings to electrons will be generated at 1-loop in the electro-weak couplings.

\subsection{Axion-like particles}
In the models under consideration one often obtains extra singlets that are not anomalous under QCD.
Their mass is only generated from explicit breaking effects of $SU(N_F)_L \times SU(N_F)_R$. 
Therefore the mass of axion-like particles is expected to be related to the explicit PQ breaking, i.e. 
the size of $\bar\theta$.

Adding the PQ breaking interaction ${\cal O}_d / M_p^{d-4}$ with ${\cal O}(1)$ the effective potential for $a$ becomes,
\begin{align}
%V \sim \frac{f_\pi^2m_\pi^2}{f_{\rm PQ}^2} a^2
%+ \frac{f_{\rm PQ}^{d-1}}{M_{\rm Pl}^{d-4}} a
%+ \frac{f_{\rm PQ}^{d-2}}{M_{\rm Pl}^{d-4}} a^2.
V(a) \sim \frac{1}{2}m_a^2 a^2
+ \frac{\langle {\cal O}_d\rangle}{M_p^{d-4}} \frac{a}{f_{\rm PQ}}
+ \frac{\langle {\cal O}_d\rangle}{M_p^{d-4}} \frac{a^2}{f_{\rm PQ}^2} + \cdots, \label{eq:pot_axion}
\end{align} 
where $m_a\sim m_\pi f_\pi/f_{\rm PQ}$ is the axion mass from non perturbative QCD contributions. 
The term linear in $a$ induces a VEV for $a$,
\begin{align}
\bar\theta = \frac{\langle a\rangle}{f_{\rm PQ}}
%\sim \frac{f_{\rm PQ}^d}{M_{\rm Pl}^{d-4}} \frac{1}{f_\pi^2 m_\pi^2}.
\sim \frac{\langle {\cal O}_d\rangle}{f_{\rm PQ}^2 m_a^2 M_p^{d-4}}.
\end{align}
that should be smaller than $10^{-10}$. The third term of eq.~(\ref{eq:pot_axion}) gives  a contribution to the axion mass,
\begin{align}
\delta V_{\rm mass}
\sim \frac{\langle {\cal O}_d\rangle}{M_p^{d-4}} \frac{a^2}{f_{\rm PQ}^2}
\sim  \bar\theta m_a^2 a^2.
\end{align}
While this contribution is negligible for the QCD axion a similar contribution is generated for 
axion-like particles (ALP), if the structure of the explicit symmetry breaking  operators is generic,
\begin{align}
m_{\rm ALP} \sim \sqrt{\bar\theta} m_a < 10^{-5}\,m_a
\label{alpmass}
\end{align}

ALPs with mass given by eq. (\ref{alpmass}) would give a negligible contribution to the dark matter density. 
Indeed the relics abundance is given by \cite{Arias:2012az},
\begin{equation}
%\frac {\rho_{ALP}}{\rho_{DM}}\approx 0.2 \sqrt{\frac{m_{ALP}}{\rm eV}}\left(\frac {f_{PQ}}{10^{11}\,\rm GeV}\right)^2 \langle \theta^2 \rangle
\frac {\rho_{ALP}}{\rho_{DM}}\approx 1.6 \sqrt{\frac{m_{ALP}}{\rm eV}}\left(\frac {f_{PQ}}{10^{11}\,\rm GeV}\right)^2 \langle \theta^2 \rangle
\end{equation}
which is subdominant to the QCD axion density in the entire parameter range.

For masses  $10^{-33}\,{\rm eV} < m_{\rm ALP}<  4 \times 10^{-28}\,  {\rm eV}$ the ALPs could give measurable effects in the
precession of the CMB.
Also, for masses  $10^{-18}\,{\rm eV} \lesssim m_{\rm ALP} \lesssim 10^{-10}\,  {\rm eV}$
the ALPs could affect rotation of blackhole,  see Ref.~\cite{Arvanitaki:2009fg} and references therein.

%
%And also, there would be mixing term between the QCD axion $a$ and axion-like particles $\tilde a$, as ${\cal L} \sim \bar\theta m_a^2 a \tilde a$.
%Then, $\tilde a$ and $a$ mix with mixing angle $\bar\theta$.
%\begin{align}
%{\cal L} \sim 
%\frac{\bar\theta}{f_{\rm PQ}} \frac{1}{32\pi^2} \tilde a G \tilde G
%+ \frac{\bar\theta}{f_{\rm PQ}} \frac{1}{32\pi^2} \tilde a F \tilde F.
%\end{align}

\subsection{Topological Defects and the scale of Inflation}

The composite axion scenario suffers in general from the presence of topological defects. 
This seems an unavoidable property of the accidental symmetries required for the axion solution  of the strong CP problem.  
Beside global strings from the spontaneous $U(1)_{PQ}$  breaking there are also domain walls and  stable hadrons. 
Presence of such stable objects  during the cosmological history of the universe is strongly constrained.

The $SU(N_c)$ theories considered so far always contain baryon states that are stable because of accidental symmetries.
The quality of the PQ symmetry suggests that the baryons are cosmologically stable. If the baryon symmetry is broken 
by dim 12 operators for example the rate scales as,
\begin{equation}
\tau \sim \frac {8\pi\,M_p^{16}}{m_B^{17}} \sim 10^{-40}~{\rm s}\times \left(\frac {M_p}{4\pi f}\right)^{17}
\end{equation}
using $M_p=2.4\times 10^{18}$ GeV and estimating the mass of the baryon as $4\pi f$.
The lifetime is longer than the age of the universe for $f< 10^{13}$ GeV.
If the stable baryons are thermalized in the early universe they would overclose the universe.  
This problem could be circumvented in theories that do not admit stable baryons such as $Sp(N)$ gauge theories
with pseudo-real representations.

The domain wall  problem is even more model independent. The existence of domain walls is associated to the anomaly coefficient
that in general break $U(1)_{PQ}$ to $Z_N$. Because of the multiplicity factor $N_c$ the domain wall number cannot be 1 and therefore some stable domain walls exist
and will be produced in the early universe. This scenario would then be ruled out unless some explicit breaking of the global symmetry
is included, which however is hard to reconcile with the required quality of the PQ symmetry.

To avoid all the problems with topological defect the simplest possibility is to assume that PQ symmetry 
is broken during/before the inflation, and the reheating temperature is much smaller than $f_{\rm PQ}$.
With this choice the initial displacement of the axion is determined anthropically by the DM abundance so that
the the whole visible patch of the universe has the same $\theta_0$. Note that $f_{PQ}$ can be larger than $10^{12}$ GeV in this case
which however exacerbates the issue of accidental symmetries.
The main constraint comes in this case from isocurvature perturbations of CMB \cite{Steinhardt:1983ia}
Using the latest results from the Planck collaboration gives a constraint on $H_{\rm inf.}$ as \cite{Ade:2015lrj},
\begin{align}
H_{\rm inf.} < 0.86 \times 10^7~{\rm GeV}\left( \frac{f_{\rm PQ}}{10^{12}~{\rm GeV}} \right)^{0.408} \qquad (95~\%~{\rm C.L.}),
\end{align}
where $H_{\rm inf.}$ is the expansion rate at Hubble radius exit of the scale corresponding to $k_{\rm mid} = 0.05~{\rm Mpc}^{-1}$.
%Tensor-to-scalar-ratio \cite{Ade:2015lrj},
%\begin{align}
%V_* = (1.88\times 10^{16}~{\rm GeV})^4 \frac{r}{0.10}.
%\end{align}
On the other hand, the energy density $V_*$ during the inflation can be written by using tensor-to-scalar ratio $r$ as
$V_* = (3.34\times 10^{16}~{\rm GeV})^4\,r $ \cite{Ade:2015lrj}.
Thus,
\begin{align}
H_{\rm inf.} =
%\sqrt{\frac{V_*}{3M_P^2}} =
2.7 \times 10^{14} \,\sqrt{r}~{\rm GeV}.
\end{align}
Of course if non-zero tensor-to-scalar ratio were observed in  future experiments,  composite axion models would be severely constrained.

\subsection{Unification}
\label{unification}

\begin{table}[t]
\begin{center}
\begin{tabular}{|c|ccc|c|c|ccc|c|c||}
\hline 
$SU(5)$ & $SU(3)_c$ & $SU(2)_L$ & $U(1)_Y$ & charge & name  & $\Delta N$ & $\Delta E$ &\\
\hline \hline
\rowcolor[cmyk]{0,0,0.2,0} $1$ & $1$ & $1$ & $0$ & $0$ & $N$   & 0 & 0 & \\  \hline
\rowcolor[cmyk]{0,0.2,0,0.1} $ \bar{5}$ & $\bar{3}$ & $1$ & $1/3$ & $1/3$ & $D$  & 1/2 & 1/3 & \\ 
\rowcolor[cmyk]{0,0.2,0,0.1}  & $1$ & $2$ & $-1/2$ & $0$,\,$-1$ & $L$  & 0 & 1 &\\
\hline
\rowcolor[cmyk]{0.1,0,0.1,0} $10 $ & $\bar{3}$ & $1$ & $-2/3$ & $-2/3$ & $U$  & 1/2 & 4/3 & \\
\rowcolor[cmyk]{0.1,0,0.1,0} & $1$ & $1$ & $1$ & $1$ & $E$   & 0 & 1 & \\
\rowcolor[cmyk]{0.1,0,0.1,0} & $3$ & $2$ & $1/6$ & $2/3$,\,$-1/3$ & $Q$  & 1 & 5/3 & \\
\hline
\rowcolor[cmyk]{0.2,0.0,0,0.0} $15 $ & $3$ & $2$ & $1/6$ & $2/3$,\,$-1/3$ & $Q$ & 1 & 5/3 &  \\
\rowcolor[cmyk]{0.2,0.0,0,0.0} & $1$ & $3$ & $1$ & $0$,\,$1$,\,$2$ & $T$ &  0 & 5 & \\
\rowcolor[cmyk]{0.2,0.0,0,0.0} & $6$ & $1$ & $-2/3$ & $-2/3$ & $S$ & 5/2  & 8/3 & \\
\hline
\rowcolor[cmyk]{0,0,0.2,0} $24$ & $1$ & $3$ & $0$ & $-1,\,0,\,1$ & $V$ & 0 & 2 & \\
\rowcolor[cmyk]{0,0,0.2,0} & $8$ & $1$ & $0$ & $0$ & $G$ & 3 & 0 & \\
\rowcolor[cmyk]{0,0,0.2,0} & $\bar{3}$ & $2$ & $5/6$ & $4/3$,\,$1/3$ & $X$ & 1 & 17/3 & \\
\rowcolor[cmyk]{0,0,0.2,0} & $1$ & $1$ & $0$ & $0$ & $N$ & 0 & 0 &  \\
\hline
\end{tabular}
\end{center}
\caption{\label{tab:unificaxion}  SM representations arising from  the smallest $SU(5)$ representations. 
We give the SM decomposition, assign standard names used throughout the paper,  and list the anomaly contributions  $\Delta N$  and $\Delta E$ 
for a chiral rotation with unit charge.}
\end{table}

The accidental axion models discussed in the previous sections can be easily extended to unified theories. 
First of all, compatibility with unification requires that the SM representation are branches of unified representations, the smallest $SU(5)$ multiplets are given in 
table \ref{tab:unificaxion}. This implies some restrictions on the possible charges, for example color triplets cannot have hypercharge equal to zero.

One possibility is to construct the theory with complete $SU(5)$ multiplets. Let us consider the vectorial models. The simplest model along these lines
is obtained by taking fermions in the $5+1$ representation that contains $D+L+N$ states. In this case the axion can be identified with $SU(5)$ invariant 
traceless combination of charges,
\begin{equation}
D+L- 5N
\end{equation}
Note that another SM singlet exists (corresponding to the rotation $2D-3L$) whose shift symmetry is explicitly broken by unification. 
Its mass is however very large, of order  $f_{PQ}^2/M_{GUT}$.
For the QCD axion a model independent prediction whenever unified multiplets are used is that,
\begin{equation}
\frac E N= \frac 8 3
\end{equation}
In this case the differential of running of SM couplings is not modified so that their unification is only approximate as in the SM,
though enhanced threshold corrections are expected.

Unification of SM couplings could be improved with incomplete multiplets as in \cite{unificaxion}.
Above the confinement scale the differential running is modified and unification of couplings 
can be achieved. We can build for example a CK model with moose structure based on the choice of fermions \cite{axionhiggs}, 
\begin{equation}
\Psi= D+ L+  Q+  U
\end{equation}
which corresponds to a 5 and an incomplete 10 rep of $SU(5)$. From the point of the differential running this is equivalent to a shift 
of $\beta-$function coefficients induced by  $- E$ multiplied by the number of colors. 
One finds that unification of couplings can be achieved for $N=6$ colors with,
\begin{equation}
\alpha_{\rm GUT}\approx 0.05\,,~~~~~~~~~~~~~~~~~~~~m_{\rm GUT}\approx 10^{17}\, {\rm GeV}
\end{equation}
assuming that the running is modified at a scale around $10^{12}$ GeV.

One significant difference compared to \cite{unificaxion} is that the relation between unification and the ratio $E/N$ is model dependent.
In that Ref. the axion was realised with a single elementary complex scalar field that spontaneously breaks the PQ symmetry. Because of renormalizability all the fermions
have the same PQ charge  so that a model independent relation between $\beta-$function coefficients and $E/N$ is found. In the composite axion models
different fermions have different PQ charge as determined by the combination that couples to the QCD anomaly. 
In the model above the axion, assuming that $D+L$ and $Q+U$ belong to a single 5 and a 10 respectively, corresponds to the combination of charges,
\begin{equation}
9(D+L) - 5(Q +U)
\end{equation}
from which we derive,
\begin{equation}
\frac E N= \frac 6 5
\end{equation}

These results can also be extended to axial models. In this case above the confinement scale the gauge group 
becomes $SU(m)_L\times SU(5)_R$ or $SO(2m)_L\times SU(5)_R$. In order to avoid exotic states the SM 
should be a representation of $SU(5)_R$.  Unification would proceed in a similar way as discussed above.

\section{Summary}
\label{sec:conclusions}

To conclude we wish to cast our results in a model independent way. The starting point to construct a composite axion model is the spontaneous breaking of $G\to  H$. 
Beside $SU(N_F)\times SU(N_F)/SU(N_F)$, UV complete models  can be obtained from QCD-like dynamics with appropriate gauging of the global symmetries \cite{thaler}. 

Among the Goldstone bosons, SM singlets are typically anomalous under QCD and electromagnetism and  thereby  provide natural axion candidates. 
The anomalies are determined by the Wess-Zumino-Witten term of the chiral Lagrangian and in particular the coupling to photons is fixed by group theory. 
In order to have anomalies $G$ must contain $SU(N)$ or $U(1)$ factors.

In order for the axion to solve the strong CP problem QCD should be the dominant source of breaking of the axion shift symmetry to exquisite precision.
This depends on the UV completion that realizes the spontaneous breaking. For vectorial theories this requires that the fermions masses are set to zero and moreover higher 
dimensional operators should also be eliminated up  to dimension at least 11. The axion can  arise accidentally if the gauge theory is chiral. Within the composite axion scenario 
this can be achieved either with a chiral gauging of the global symmetry of the coset or introducing a moose structure with nearest neighbour interactions. 
The latter case is equivalent to the deconstruction of a 5-dimensional gauge theory. Suppression of the interactions is then  understood  from locality. 

Composite axion scenarios most likely require the scale of inflation to be smaller than $f_{PQ}$ in order to avoid severe constraints from the existence of
topological defects.  It is of course extremely difficult if not impossible to determine whether the axion is composite from its low energy properties. 
Generic phenomenological features are however additional axion-like-particles and model independent derivative couplings to matter. 
The composite axion demands  the existence of new fermions with SM charges that could improve the unification of SM couplings. 
Knowledge of the electromagnetic coupling to photons would provide a precious information to select the relevant models.
 
\section*{Acknowledgements}
The work of MR is supported in part by the MIUR-FIRB grant RBFR12H1MW.
The work of RS is supported in part by JSPS Research Fellowships for Young Scientists. We would like to thank Ryuichiro Kitano for
discussions. MR would like to thank Giovanni  Villadoro for many conversations on axions and for comments on the manuscript.

\appendix

\section{$SO(N)$ models}
\label{app}

The dynamics of $SO(N_c)$ gauge theories with fermions in a vector representation presents some new features.
Due to the reality of the representation the pattern of symmetry breaking is expected to be,
\begin{equation}
\frac{SU(N_F)}{SO(N_F)}
\end{equation}
determined by the vacuum condensate,
\begin{equation}
\langle \psi^i \psi^j\rangle= \Lambda^3 \delta^{ij}
\end{equation} 
Consider an $SO(N_c)\times SU(M)\times SU(M)$ gauge theory with Weyl fermions in the chiral rep,
\begin{equation}
(N_c,m, \bar{m})+ m (N_c,\bar{m},1)+ m (N_c,1, m)
\end{equation}
The global symmetry of $SO(N_c)$ is $SU(3 m^2)$. The dynamics of the theory depends on the vacuum alignment since both $SU(M)$ gauge symmetries cannot 
be preserved simultaneously. The largest unbroken group is the diagonal combination of $SU(M)$
so we will assume that the dynamics does not break it. Consider now the $U(1)$ symmetries that rotate the representations above with charges $q_1$, $q_2$ and $q_3$.
The anomalies read,
\begin{eqnarray}
%&&U(1) \times  SU(N_c)^2: ~~~~~~~~~~~2 m^2 (q_1+q_2+q_3)\nonumber \\
&&U(1) \times  SO(N_c)^2: ~~~~~~~~~~~2 m^2 (q_1+q_2+q_3)\nonumber \\
&&U(1) \times  SU(M)_1^2: ~~~~~~~~~~~2 m N_c (q_1+q_2) \nonumber \\
&&U(1) \times  SU(M)_2^2: ~~~~~~~~~~~2 m N_c (q_1+q_3)
\end{eqnarray}
Since these 3 combinations are linearly independent, as in $SU(N_c)$ examples all 3 $\theta$ angles can be removed,
independently of the vacuum alignment.

The strong CP problem is solved similarly to section \ref{sec:axialgauge}. The condensates are, 
\begin{equation}
r_1 r_1\,,~~~~~~~~~~~~r_2 r_3
\end{equation}
%The vacuum does not break the following generator
%\begin{equation}
%T_1=(0,1,-1):~~~~~~~~~~~~~~A_1= 2 m N_c (0,1,-1)
%\end{equation}
The spontaneously broken $U(1)$ generators and their anomaly coefficients under $SO(N_c)\times SU(M)\times SU(M)$ are,
\begin{eqnarray}
&T_2=(0,1,1):~~~~~~~~~~~~~~A_2=(4m^2,2 m N_c,2 m N_c) \nonumber \\
&T_3=(1,0,0):~~~~~~~~~~~~~~A_3=(2 m^2,2 m N_c,2 m N_c)
\end{eqnarray}
The effective Lagrangian for the axions contains,
\begin{eqnarray}
{\cal L}_{eff}&\sim& \frac {a_1}{f_{PQ}}( 4 m^2 F_1^2+2 m N_c F_2^2+ 2 m N_c F_3^2)+\frac{a_2}{f_{PQ}} (2 m^2 F_1^2+2  m N_c F_2^2+2  m N_c F_3^2)\nonumber \\
&=& 2 m^2\frac{2a_1+ a_2}{f_{PQ}} F_1^2 + 2 m N_c \frac{a_1+a_2}{f_{PQ}} F_g^2  
\end{eqnarray}
so that the physical axion can be identified with the combination,
\begin{equation}
a_{QCD}\sim \frac {a_1-2 a_2}{\sqrt{5}}
\end{equation}
that corresponds to the generator $T_{QCD}=1/\sqrt{5} (-2,1,1)$.

The constraints on asymptotic freedom are in this case as follows,
\begin{equation}
b_{N}= - \frac {11}3 (N_c-2) + \frac 2 3 N_F~~~~~~ \rightarrow~~~~~~~ m^2<\frac {11} 8 (N_c-2)
\end{equation}
For the running of $SU(M)$ we have,
\begin{equation}
\frac 1 {\alpha_s(M)}\approx30+ \frac 7 {2\pi} \log \frac M {10^{12}\,{\rm GeV}}- \frac {m N_c} {3\pi} \log \frac M {10^{12}\,{\rm GeV}} 
\end{equation}
This is identical to what one obtains in the model of section \ref{sec:randall}.

In this model PQ symmetry can be violated by dimension 6 operators.
This can again be  improved by adding a moose structure to the theory with $SU(N_c)$ gauge fields and gauging diagonal $SU(M)$ global symmetries.

\end{document}